\documentclass[iop,apj]{emulateapj}
\begin{document}
\newcommand{\beq}{\begin{equation}}
\newcommand{\eeq}{\end{equation}}
\def\eeqno#1{\label{#1}\end{equation}}
\def\div{\vec\nabla\cdot}
\def\grad{\vec\nabla}
\def\curl{\vec\nabla\times}
\def\rar{\rightarrow}
\def\re{R_{e}}
\def\az{a_0}
\def\cmst{~{\rm cm~ s}^{-2}~}
\def\gcmt{~{\rm g~ cm}^{-2}~}
\def\vinf{V_{\infty}}
\def\Halpha{H$_{\alpha}~$}
\def\Sz{\Sigma_0}
\def\s{\sigma}
\def\a{\alpha}
\def\b{\beta}
\def\l{\lambda}
\def\deg{^o}
\def\kpc{{\rm kpc}}
\def\mpc{{\rm Mpc}}
\def\gpc{{\rm Gpc}}
\def\St{\Sigma_t}
\def\vg{{\bf g}}
\def\vh{{\bf h}}
\def\vr{{\bf r}}
\def\vR{{\bf R}}
\def\_#1{_{\scriptscriptstyle #1}}
\newcommand{\mss}{\ensuremath{\mathrm{m}\,\mathrm{s}^{-2}}}
\newcommand{\kms}{\ensuremath{\mathrm{km}\,\mathrm{s}^{-1}}}
\newcommand{\MLsun}{\ensuremath{\mathrm{M}_{\sun}/\mathrm{L}_{\sun}}}
\newcommand{\Lsun}{\ensuremath{\mathrm{L}_{\sun}}}
\newcommand{\Msun}{\ensuremath{\mathrm{M}_{\sun}}}
\newcommand{\Aunits}{\ensuremath{\mathrm{M}_{\sun}\,\mathrm{km}^{-4}\,\mathrm{s}^{4}}}
\newcommand{\surfdens}{\ensuremath{\mathrm{M}_{\sun}\,\mathrm{pc}^{-2}}}
\newcommand{\etal}{et al.}
\newcommand{\LCDM}{$\Lambda$CDM}
\newcommand{\ML}{\ensuremath{\Upsilon_*}}

\title{Andromeda Dwarfs in Light of MOND}
\author{Stacy  McGaugh}
\affil{Department of Astronomy, Case Western Reserve University, Cleveland, OH 44106, USA}

\and

\author{Mordehai Milgrom}
\affil{Department of Particle Physics and Astrophysics, Weizmann Institute of Science, Rehovot 76100, Israel}

\begin{abstract}
We compare the recently published velocity dispersions for seventeen Andromeda dwarf spheroidals with estimates  of the MOND predictions,
based on the luminosities of these dwarfs, with reasonable stellar $M/L$ values, and no dark matter.
We find that the two are consistent within the uncertainties.  We further predict the velocity dispersions of another ten dwarfs
for which only photometric data are currently available.
\end{abstract}


\section{Introduction}

The Modified Newtonian Dynamics (MOND) \citep{milg83} is a hypothesis that explains the mass discrepancies observed
in galaxies without dark matter.  It has recently been reviewed by \citet{famaey12}.
In MOND, the dynamics is conventional at large accelerations but modified at low accelerations $a \lesssim a_0 = 1.2 \times 10^{-10}\;\mss$.
In the limit of very low accelerations, it is hypothesized that the acceleration is enhanced over the usual Newtonian
value $g_N$, tending towards a deep MOND limit $a \rightarrow \sqrt{g_N a_0}$ for
$a \ll a_0$.\footnote{This formulation occurs naturally if dynamics is required to be invariant under space-time scaling, $(t,\bf{r})\rightarrow \lambda(t,\bf{r})$.
In self-gravitating systems, the Newtonian force per unit mass $g_N\sim MG/r^2$ scales as $\lambda^{-2}$,
while accelerations, $a$, scale as $\lambda^{-1}$ ($M$ is the observed luminous mass). We can then not have $a=g_N$, which is not scale invariant.
Instead, a relation of the form $a\sim \sqrt{a_0 g_N}$ is implied. For any isolated mass we must have asymptotically far $a=q\sqrt{a_0 g_N}$,
with a universal constant $q$; $a_0$ is normalized so that $q=1$ \citep{milgrom12}}

The predictions of MOND have largely been confirmed in rotationally supported galaxies \citep[for reviews and reference see][]{SM02,famaey12}.
It has been less extensively tested in pressure supported systems, in part because bright elliptical galaxies only reach the MOND regime at
large radii where the data become sparse \citep[see, however,][]{dearth,Richtler11,milgromEll}.  The predictions of the theory become most acute
in systems of low surface brightness, where the characteristic baryonic surface density  $\Sigma_b \ll a_0/2\pi G$, which
corresponds to $g_N\ll a_0$, and hence to $a\ll a_0$.

The dwarf Spheroidal (dSph) satellites of the Milky Way and Andromeda provide a set of very low surface density, pressure supported
stellar systems in which the predictions of MOND can be tested.  This test was first applied to the classical dwarfs of the Milky Way
by \citet{GS92}.  They found that while five of the seven dwarfs were in good agreement with MOND, two (Draco and Ursa Minor) required
unpleasantly high mass-to-light ratios.  These improved as the data improved \citep{milg7dw}, but persist in being marginal cases
when analyzed with only a global velocity dispersion \citep{mw10}.
One must always bear in mind the applicability of the necessary assumptions; in Ursa Minor for example there is evidence for multiple
dynamical components \citep{UMIsubstruct} which, if true, would considerably complicate the analysis. However, the more basic issue was whether we should be more impressed by the five cases that are in good agreement with MOND or the two that disagree.

The data for the classical Milky Way dwarfs are considerably better now.  \citet{walker07} have measured the velocity dispersion profile as a function of radius.
This enables detailed analysis of their mass profiles, not just a single global characteristic value.
\citet{angusdw} and \citet{serradw} found these to be in good accord with MOND provided that the orbital anisotropy varies with radius,
much as found in conventional analyses \citep{anisotropy}.  At this level of detail, one must also be cognizant of tidally liberated interlopers
\citep{serradw}, and differing velocity dispersion profiles for different stellar populations \citep[e.g.,][]{batt06,batt08}.

In addition to improved data, the available sample has grown rapidly with the discovery of many new dwarf satellites around both the Milky
Way and Andromeda.  The so-called ultrafaint dwarfs are discussed by \citet{mw10}, who find that these tiny galaxies have velocity dispersions
larger than expected for isolated systems  of their mass in MOND. However, these are not isolated systems.  Tidal effects due to the
Milky Way cannot be ignored, obviating the usual assumption of dynamical equilibrium.  It will be interesting to see if
any of the ultrafaints are in the process of dissolution anticipated by \citet{bradadwarf}.

An early test of MOND with the dwarfs of M31 was made by \citet{andiicote}, who analyzed And II.  For a measured velocity dispersion of
$\s=9.3\;\kms$, they found good agreement with MOND with no dark matter: $M_*/L_V=3.2\;\MLsun$ instead of $M/L_V=21\;\MLsun$
in the Newtonian analysis.  Differing velocity dispersions have since been published, and the estimated luminosity of And II has tripled
\citep{kalirai10,tollerud2012}, lowering $M*/L$ accordingly.
We extend the MOND analysis to the many more recently measured dwarfs (and updated data for And II).

There are at present seventeen Andromeda dwarfs with published velocity dispersion measurements that can be employed for this test.
There are a further ten that do not yet have published velocity dispersions
but for which there exists sufficient photometric information that a prediction can be made.

These dwarfs all exhibit large mass discrepancies, requiring most of their mass to be dark in Newtonian dynamics. On the other hand, the recent realization
\citep{ibata13,conn13} that many (15) dwarfs (including twelve that we analyze here) are part of a large, kinematically coherent disc of satellites,
is at odds with the notion that they are the primordial occupants of the sub-halos found in \LCDM\ simulations:  their phase space distribution is quite
different.  Sub-halos comprise a dynamically hot, quasi-spheircal population rather than a thin, rotating disk.
Arguments to the same effect have been made in connection with a similar disk of satellites around the Milky Way \citep{kroupafalse,pawlowski12}.
\citet{dabringhausen13} suggest that these orbital properties are the result of formation as tidal dwarfs.
Tidal dwarfs should not exhibit the observed mass discrepancies in \LCDM\ because of the difference in phase space between dark matter and baryonic
tidal debris \citep{bournaud}, but this situation can arise naturally in MOND \citep{milgrom2007,gentiletidal,tiret}.

In MOND we predict at the outset that these dwarfs should show large mass discrepancies, as observed.
This is based on the  strong correlation between low acceleration and high mass discrepancy predicted by MOND \citep{milg83}.
These low surface brightness systems all have very low internal accelerations.
The acceleration at the half mass radius is estimated to be between $0.01a_0$ and $0.15 a_0$ for all dwarfs considered here,
with only three exceeding $0.1a_0$.

In \S \ref{estimates}, we describe the MOND estimators we use for the bulk velocity dispersions.
In \S \ref{data} we discuss the various data we use.  In \S \ref{results}, we describe our results.
These include a comparison of the observed velocity dispersions with those predicted by MOND for seventeen dwarfs (\S \ref{indcase}).
For another ten dwarfs, we predict the velocity dispersion in advance of measurement (\S \ref{furtherpredictions}).
We conclude in \S \ref{discussion}.

\section{MOND Estimates of the Velocity Dispersion}
\label{estimates}

Our basic estimate of the predicted MOND line-of-sight velocity dispersion is based on the deep-MOND virial relation
 \beq MG\az=\frac{9}{4}\langle v^2\rangle^2.\eeqno{i}
Here, $\langle v^2\rangle$ is the 3-D mass weighted mean-square velocity, and
$M$ is the total mass of the system. This relation holds exactly in the deep-MOND limit of the modified Poisson equation of \citet{bm84} for isolated,
self-gravitating, steady-state (virialized) systems made of test particles\footnote{Namely, constituents whose masses are all
much smaller than the total mass of the system.
There is a simple generalization of this relation to systems made of arbitrary masses \citep{milgrom97,milgrom10}.} \citep[see][]{milgrom94,milgrom97}.
It is also exact in the QUMOND formulation of MOND, under the same conditions \citep{milgrom10}.
More generally, for any isolated, deep-MOND, self-gravitating system, a relation of the form $Q^2 MGa_0=V^4$ should hold for any
MOND theory \citep[as it follows from the basic assumptions of MOND:][]{milgrom12}, where $V$ is some measure of the mean 3-D velocity,
and $Q\sim 1$ may depend on the exact theory, and on details of the system.

The available data on the M31 dwarfs give only some measure the line-of-sight velocity dispersion, $\s$, which we want to compare
with $\langle v^2\rangle^{1/2}$ deduced using eq.\ (\ref{i}). For the comparison to be meaningful we have to assume, in default of better knowledge,
that the M31 dwarfs are globally isotropic, namely that $\s$ would be independent of the line of sight.  This would not be quite correct if rotation is
present, or if the system is otherwise globally anisotropic. Furthermore, we assume that the quoted $\s$ represents the mass-weighted dispersion
over the whole system, which may not be quite correct in general. Specifically, in the case of the M31 dwarfs $\s$ is based only on measured velocities
of red giants, and the spatial coverage of each system is not always complete.  It is not clear to what extent the necessary assumption that
the observed velocity dispersion of a particular stellar population along our line of sight is a valid approximation of the desired 3D mass weighted
velocity dispersion.

Making all those assumptions permits us to represent  $\langle v^2\rangle$ by $3\s^2$, and thus to write the line-of-sight
dispersion predicted by MOND as \beq \s_{iso} \approx (\frac{4}{81}MG\az)^{1/4}.\eeqno{ii}
We use the subscript \textit{iso} to specify the case of an isolated object not influenced by external fields.

The assumption of isolation is clearly violated for some of the M31 dwarfs.
Isolation here means that the system is not subject to an appreciable external-field effect (EFE) \citep{milg83}.
Let $g_{in}$ represent accelerations within the system (here, the dwarf) with respect to its center of mass, and
$g_{ex}$ the acceleration\footnote{Here $g_{in}$ is taken as the acceleration on a star at the half-mass radius due to the mass of the dwarf itself,
while $g_{ex}$ is the acceleration on the center-of-mass of the dwarf due to M31.  These are the actual accelerations, not just the
Newtonian expectation $g_N$ for the baryons alone.}
with which the system falls as a whole --- here the acceleration produced by M31 at the position of the dwarf.
Then, relation (\ref{ii}) applies only if $g_{in}\gg g_{ex}$. Some of the M31 dwarfs satisfy this condition to a reasonable approximation.
In the opposite case, $g_{in}\ll g_{ex}$, the EFE is dominant. In such cases, the above mentioned MOND theories predict that the internal
dynamics of the system is quasi-Newtonian: the Newtonian virial relation is satisfied with two corrections: i) instead of Newton's constant $G$
one has to use a larger effective value $G_{eff}=G/\mu(g_{ex}/\az)$, where $\mu$ is the MOND interpolating function, and ii) the dynamics is
not rotationally invariant, since the external field introduces a preferred direction. We thus expect some induced anisotropy.
In the case of EFE dominance, our estimator becomes
\beq \s_{efe} \approx (\frac{MG_{eff}}{3 r_{1/2}})^{1/2}, \eeqno{iii}
where $r_{1/2}$ is the 3D half light radius of the dwarf.

It so happens, however, that the M31 dwarfs that are not relatively free of the EFE, are also not of the other extreme type: they
have $g_{in}\approx g_{ex}$. It is then rather difficult to make an estimate of the MOND prediction for $\s$. For these cases we simply give
the predictions MOND would make for $\s$, for both the isolated case according to eq.\ (\ref{ii}), and the EFE dominance case of eq.\ (\ref{iii}).
The predicted value of $\s$ for a given mass should always be smaller when the EFE is important, provided the system remains in
dynamical equilibrium.  Systems subject to very strong external fields may suffer tidal stirring and disruption.

For each dwarf we estimate $g_{ex}=V^2_{M31}/D_{M31}$, where $V_{M31} \approx 230\;\kms$ \citep{M31RC} is an estimate of the asymptotic
rotational speed of M31\footnote{The predicted EFE velocity dispersion $\s_{efe}$ depends inversely on $V_{M31}$, so would be 10\% lower if $V_{M31}$
were 10\% high ($253\;\kms$).  The approximation is necessarily crude, as for some of the more distant dwarfs, the acceleration due to other
objects may begin to play a role (e.g., the effect of M33 upon And XXII, where the opposing vectors of the larger galaxies may tend to cancel
one another and reduce the EFE).},
and $D_{M31}$ is the distance from the dwarf to the center of M31.
We estimate $g_{in}=3\s^2/r_{1/2}$, where $\s$ is calculated from eq.\ (\ref{ii}).
These are used to determine the importance of the EFE,
and $g_{ex}$ is also used in calculating $G_{eff}$.  In the limit of very small accelerations that is satisfied here,
$G_{eff} \approx G\az/g_{ex}$.

\section{The data}
\label{data}

Spectra of individual stars in dwarf spheroidal galaxies are used to construct the velocity dispersions of their stellar population.
At the distance of M31 ($\sim 780$ kpc), only red giants are amenable to observation.  It is possible for different stellar populations
to have different velocity dispersions; this situation is observed in Fornax \citep{batt06} and Sculptor \citep{batt08}.
In order to perform the analysis here, we make the usual assumptions that the observed, line-of-sight
velocity dispersion of the red giants is representative of the full mass-weighted 3D velocity dispersion and that the system is in dynamical equilibrium.

The assumption of dynamical equilibrium is not as safe in MOND as it is in the context of dark matter.  In the latter case, the stellar extent of
dwarf spheroidals is usually well within their inferred tidal radii:  the stars are safely ensconced within their dark-matter halos.
Tidal effects are relatively stronger in MOND\footnote{Ursa Minor, Draco, and nearly all of the ultrafaint dwarfs have photometric tidal radii
comparable to their MOND tidal radii \citep{mw10}.}, and there is no protective cocoon of dark matter.
To complicate matters, a dwarf with an eccentric orbit that brings it close to its host may suffer non-adiabatic perturbations at
pericenter \citep{bradadwarf} that may persist even as it spends most of its time near apocenter\footnote{Perhaps the tidal debris
recently observed around Carina by \citet{batt12} is an example of this effect.}.  We do not know the orbits of the dwarfs of Andromeda, only
their present distances from M31.  At their present locations, only And IX appears to be in imminent danger from tides as it has
$M/r_{1/2}^3 < M_{M31}/D_{M31}^3$ (where $M$ is the stellar mass of the dwarf and $M_{M31}$ the baryonic mass of the host).
The rest of the dwarfs of M31 are sufficiently far from their host that they should not suffer tidal disruption unless they are on rather eccentric orbits.

The luminosities, half light radii, velocity dispersions, and M31-centric distances employed here come from a variety of sources.
The most extensive kinematic survey to date is the SPLASH survey \citep{kalirai10,tollerud2012}.
In addition to providing a large and fairly homogeneous set of data, dwarfs for which there are multiple analyses illustrate how the
deduced velocity dispersion can change with further observation or simply different analysis.  Usually the results are consistent, but in
some cases the velocity dispersion changes by more than the formal uncertainties would predict (compare, for example, published estimates of
the velocity dispersions of And II, And III, and And VII).

Further data are provided by \citet{collins10,collins11} and \citet{chapman12}.  In addition to providing measurements
of dwarfs not yet covered by SPLASH, these authors also provide completely independent measurements of some of the same dwarfs,
providing another consistency check.  There are at present four cases of dwarfs with overlapping, independent data.
In the case of And V, the measured velocity dispersions are consistent.  In And IX, the velocity dispersions differ
substantially: $\sigma = 10.9\;\kms$ vs.\ $4.5\;\kms$.  In the case of And XIII there is also a difference in the measured value, but the
formal significance of this difference is less. \citet{chapman12} estimate an upper limit for the velocity dispersion of And XXII that is
consistent with the value measured by \citet{tollerud2012}.  Differences between independent studies, when present, may be
attributable to different observational techniques (e.g., fibers vs.\ slits), but do not strike us as beyond the realm of happenstance
given our experience with the dwarf satellites of the Milky Way.

\begin{deluxetable*}{lcccccccl}
\tablewidth{0pt}
\tablecaption{Predicted and Measured Velocity Dispersions}
\tablehead{
\colhead{Dwarf} & \colhead{$L_V$}  &\colhead{$r_{1/2}$}
& \colhead{$\sigma_{obs,1}$} & \colhead{$\sigma_{obs,2}$}
 & \colhead{$\sigma_{iso}$} & \colhead{$\sigma_{efe}$}
 & \colhead{$g_{in}$} & \colhead{Ref.} \\
 & \colhead{$10^5\;L_{\sun}$} & pc & \multicolumn{4}{c}{\kms} &\colhead{$a_0$} &
}
\startdata
And I     & \phn45.\phn\phn & \phn832 	&$10.2\pm1.9$         & $10.6\pm1.1$         	& $9.2_{-1.5}^{+1.7}$ & $8.0^{+3.3}_{-2.3}$ 	& 0.082 	 & 1,2 \\
And II\tablenotemark{a}    & 93.\phn         & 1660    	& $7.3\pm0.8$         & $10.0\pm1.7$         	& $11.0_{-1.8}^{+2.1}$ & \dots       	& 0.059    	 & 2,3 \\
And III   & \phn10.\phn\phn & \phn525 	& $9.3\pm1.4$         & $4.7\pm1.8$          	 &  $6.3_{-1.0}^{+1.2}$ & $5.4^{+2.3}_{-1.6}$ 	& 0.062    	 & 1,2 \\
And V     & \phn5.9\phn     & \phn292 	&$10.5\pm1.1$         & $11.5^{+5.3}_{-4.4}$ 	&  $5.5_{-0.9}^{+1.0}$ & $7.0^{+2.9}_{-2.0}$       	 & 0.085    	 & 1,4 \\
And VI    & \phn34.\phn\phn & \phn440 	& $9.4^{+3.2}_{-2.4}$ & \dots                	 & $8.6_{-1.4}^{+1.6}$  & \dots       	& 0.14\phn 	& 4 \\
And VII   & 178.\phn\phn    & \phn977 	&$13.0\pm1.0$         & $9.7\pm1.6$          	 & $13.0_{-2.1}^{+2.5}$ & \dots       	& 0.14\phn 	& 1,2 \\
And IX\tablenotemark{b}    & \phn1.5\phn     & \phn552 	 &$10.9\pm2.0$   & $4.5^{+3.6}_{-3.4}$  & $3.9_{-0.6}^{+0.7}$  & $1.5^{+0.6}_{-0.4}$ & 0.023  & 1,5 \\
And X     & \phn0.76        & \phn309 	& $6.4\pm1.4$         & $3.9\pm1.2$          	 & $3.3_{-0.5}^{+0.6}$  & $2.3^{+1.0}_{-0.7}$ 	& 0.029    	 & 1,2 \\
And XI    & \phn0.49        & \phn145 	& $\le 4.6$           & \dots                	 & $3.0_{-0.5}^{+0.6}$  & $3.2^{+1.3}_{-0.9}$      	 & 0.049    	& 5 \\
And XII   & \phn0.31        & \phn289 	& $2.6^{+5.1}_{-2.6}$ & \dots                	 & $2.6_{-0.4}^{+0.5}$  & $1.7^{+0.7}_{-0.5}$ 	& 0.020    	& 5 \\
And XIII  & \phn0.41        & \phn203 	& $5.8\pm2.0$         & $9.7^{+8.9}_{-4.5}$  	& $2.8_{-0.4}^{+0.5}$  & $2.5^{+1.0}_{-0.7}$ 	& 0.032    	 & 1,5 \\
And XIV   & \phn2.1\phn     & \phn537 	& $5.3\pm1.0$         & $5.4\pm1.3$          	 & $4.3_{-0.7}^{+0.8}$  & $3.6^{+1.5}_{-1.0}$ 	& 0.027    	 & 1,2 \\
And XV    & \phn7.1\phn     & \phn355 	& $4.0\pm1.4$         & \dots                	 & $5.8_{-0.9}^{+1.1}$  & $6.1^{+2.5}_{-1.8}$       	 & 0.076    	& 1 \\
And XVI   & \phn4.1\phn     & \phn178 	& $3.8\pm2.9$         & \dots                	 & $5.0_{-0.8}^{+1.0}$  & \dots       	& 0.12\phn 	& 1 \\
And XVIII & \phn6.3\phn     & \phn417 	& $9.7\pm2.3$         & \dots                	 & $5.6_{-0.9}^{+1.1}$  & \dots       	& 0.061    	& 1 \\
And XXI   & \phn4.6\phn     & 1023    	& $7.2\pm5.5$         & \dots                	 & $5.2_{-0.8}^{+1.0}$  & $3.7^{+1.5}_{-1.1}$ 	& 0.021    	& 1 \\
And XXII\tablenotemark{c}  & \phn0.35        & \phn340 	& $3.5_{-2.5}^{+4.2}$  & $\le 6.0$  & $2.7_{-0.4}^{+0.5}$  & $2.3^{+1.0}_{-0.7}$ & 0.018   & 1,6
\enddata
\tablerefs{The luminosities, half light radii, and velocity dispersions are taken from
1.~\citet{tollerud2012},
2.~\citet{kalirai10},
3.~\citet{Ho2012},
4.~\citet{collins11},
5.~\citet{collins10}, and
6.~\citet{chapman12}.
}
\tablecomments{The velocity dispersion predicted by MOND in the isolated and EFE limits is given for $M_*/L = 2^{+2}_{-1}\;\MLsun$.
The correct regime depends on $M_*/L$; one should generally take the lower of the two predicted velocity dispersions, so there EFE
column is omitted when it exceeds the isolated prediction for the range of considered mass-to-light ratios.
The internal acceleration at the 3D half mass radius is estimated in units of $a_0$ assuming the isolated case with $M_*/L = 2\;\MLsun$.
}
\tablenotetext{a}{And II is unusual for dSph galaxies in having measured rotation (Ho et al.\ 2012).  This is not included in
the older $\sigma_{obs,1}$,
but is included in $\sigma_{obs,2}$ through $\sigma_{obs,2} = \sqrt{\sigma^2+\onethird v^2}$.}
\tablenotetext{b}{The predictions for And IX are very uncertain as this dwarf is subject to a strong external field from M31.
This makes the assumptions of isotropy and dynamical equilibrium suspect.}
\tablenotetext{c}{We consider only the EFE due to M31, where we use $r_{1/2}$ from \citet{chapman12}. The larger radius given by \citet{tollerud2012}
would give $\sigma_{efe}$ smaller by a factor $0.8$. \citet{chapman12} argue that And XXII might be a satellite of M33 rather than M31.
The accelerations due to M31 and M33 acting on And XXII are quite uncertain, but appear to be approximately equal in magnitude yet opposite in direction.
Vector cancelation of these external fields might have the net effect of reducing the importance of the EFE.}
\label{comptable}
\end{deluxetable*}

Table~\ref{comptable} compiles the relevant observational data.  Column 1 gives the name of the dwarf.  There are gaps in the
numerical sequence when data do not exist or the relevant object turns out to be a background galaxy rather than
a satellite of M31 \citep[e.g., And IV:][]{AndIV}.
The second column is the luminosity of each dwarf and the third its 3D half light radius.\footnote{The 3D half light radius has typically been deprojected
from the observed half light radius ($r_{1/2} = 4 R_e/3$) assuming spherical systems with velocity dispersions independent of radius \citep{boom}.
The latter assumption seems reasonable given the example of the classical dwarfs \citep{walker07}, but is not guaranteed to hold.
The assumption of sphericity is a more subtle issue.  The quoted half light radii are typically major axis values, appropriate for oblate systems.
For prolate or triaxial system, the geometric mean of the axis ratios might be a better proxy for $r_{1/2}$.  Since the half-mass radius only enters the
MOND calculation for the EFE case, and then only as the square root, we simply adopt the published major axis values for consistency with
previous work.  In the most flattened case (And XXII with $\epsilon = 0.56$), adopting the smaller geometric mean for the radius would result in
a $\sim 7\%$ increase in $\s_{efe}$.  At this level, anisotropy is also likely to be important, and cause the observed line-of-sight velocity dispersion to
deviate from the predicted mass-weighted 3D velocity dispersion.}  The fourth and fifth columns give different measurements
of the observed velocity dispersion.  Examination of these two columns permit one to judge the consistency of repeat measurements.
The last column gives the references from which these data were taken.  The remaining columns present the predictions of MOND as
discussed in the next section.

\section{Results}
\label{results}

\begin{figure*}
\epsscale{1.0}
\plotone{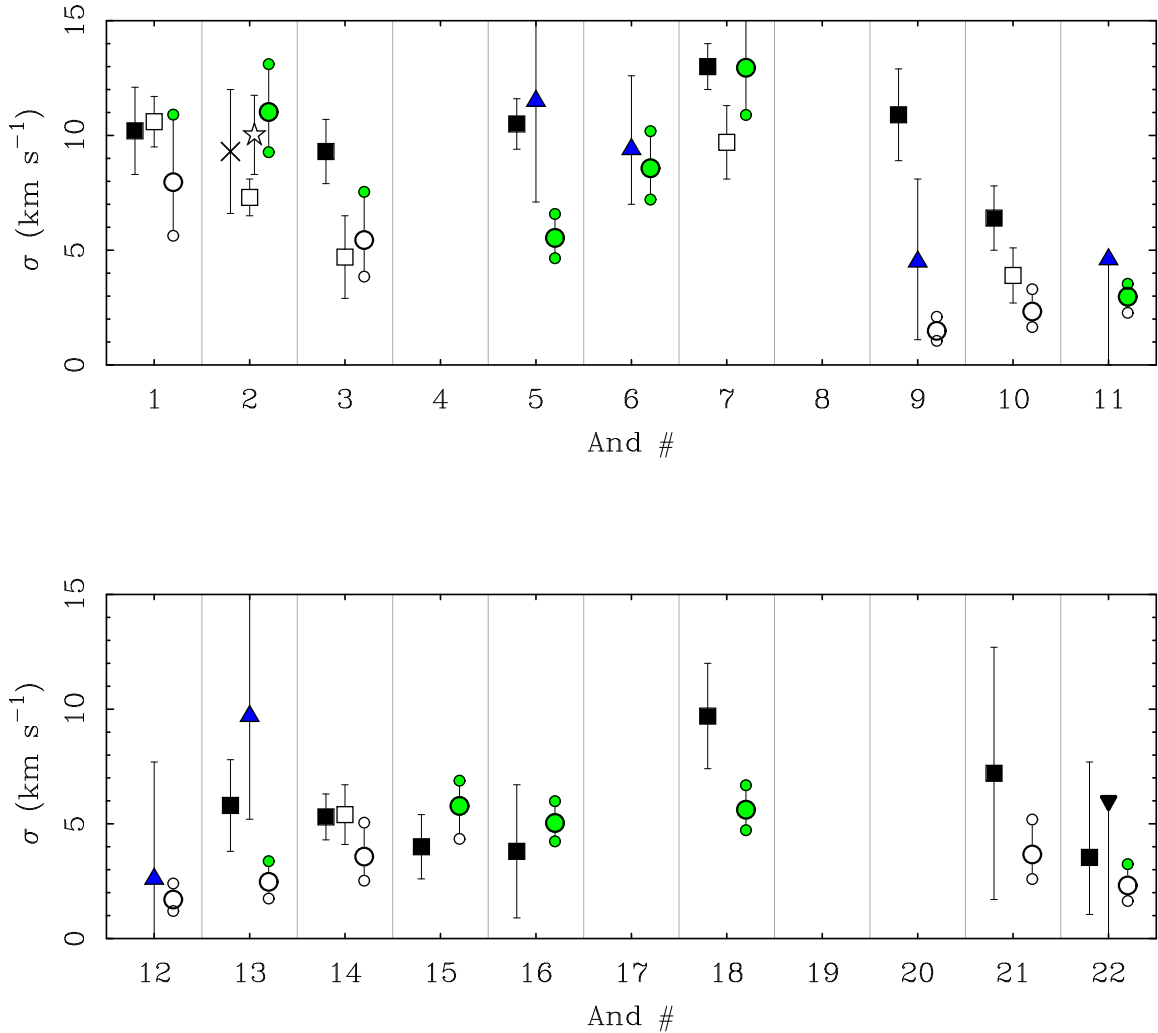}
\caption{The predicted and observed velocity dispersions of the dSph satellites of M31.  Each bin along the x-axis represents the corresponding dwarf:
And I is 1, And II is 2, and so on.  Round symbols are the predictions of MOND.  Large symbols assume $M_*/L_V = 2\;\Msun/\Lsun$.
Small symbols represent $M_*/L_V = 1\;\Msun/\Lsun$ at the lower end of the predicted range and $M_*/L_V = 4\;\Msun/\Lsun$ at the upper end.
The calculation in the isolated case is shown as a filled green circle; that when the EFE dominates is shown as an open circle.
The appropriate case is shown for each mass-to-light ratio based on which predicts the lower velocity dispersion, as discussed in the text.
Data from the SPLASH survey are shown as squares: open squares are from \citet{kalirai10}; filled squares are from \citet{tollerud2012}.
Triangles represent the data of \citet{collins10,collins11}.  In the case of And II, the early measurement of \citet{andiicote}
is shown as an $\times$, while the recent detection of rotation in this case is included in the measurement of \citet[star]{Ho2012}.
The recent upper limit in the case of And XXII \citep{chapman12}
is shown as the downward pointing triangle.}
\label{fig1}
\end{figure*}

We use the formalism described in \S~\ref{estimates} to calculate the predicted velocity dispersion for each of the dwarf spheroidals of M31,
using the photometric data\footnote{Where published photometric data differ, we adopt the non-SPLASH values as a hedge against the systematic
uncertainties that might afflict any one sample.  Affected objects are And V, IX, XIII, and XXII.  The differences are minor for all but And XXII,
where $r_{1/2}$ differs by a factor of $\sim 2$ and the luminosity differs by $\sim 50\%$.} in Table~\ref{comptable}.
The results are given in Table~\ref{comptable}, and shown in Fig.~\ref{fig1}, together with the different measured velocity dispersions.
For isolated objects in the deep MOND regime, only the stellar mass is required.  For objects where the external field dominates,
the half-mass radius is also necessary, appearing as usual in the conventional virial theorem with only the effective value
of $G$ altered (\S~\ref{estimates}).  We adopt the specific value\footnote{The value of $a_0$ obtained by \citet{BBS} remains consistent
with more recent estimates \citep{M11}.}
$a_0 = 1.21 \times 10^{-10}\;\mss$ \citep{BBS} to perform this calculation so that $(\s/\kms) = (M/1264\;\Msun)^{1/4}$.

The luminosity is typically more accurately measured than the velocity dispersion, so we use this to predict
the velocity dispersion with MOND for a range of assumed mass-to-light ratios [$M = (M_*/L)L$].
We compute $\s$ for $M_*/L_V = 1$, 2, and $4\;\MLsun$ to show the sensitivity of the result to the assumed mass-to-light ratio.
There is no guarantee that the stellar population of any given dwarf is within these limits, but it does span a plausible range.
It also gives an impression of the range in the prediction and its dependence on the stellar mass-to-light ratio.

The first calculation assumes that each dwarf is isolated.
The isolated MOND prediction $\s_{iso}$ is tabulated in column 6 of Table~\ref{comptable} where it can be compared directly to the observations.
The uncertainty in the prediction only represents the above range of mass-to-light ratios.  It does not represent the uncertainty in
any observed quantity nor does it include any of the potential systematics discussed above.
Here the luminosity is the only observational input, for which the uncertainty should be less than that in $M_*/L$.
We simply tabulate $\s_{iso}$ for $M_*/L_V = 2\;\MLsun$ and give the range for $M_*/L_V = 1$ to $4\;\MLsun$ as the uncertainty.

We confirm that the dwarfs are indeed in the MOND regime by estimating the internal acceleration
at the half light radius, again assuming $M_*/L_V = 2\;\MLsun$.  This is given in column 8 of Table~\ref{comptable} in units of $a_0$.
All of the dwarfs considered here have $g_{in} < 0.15 \az$, and often considerably less, as expected for systems of low
surface brightness\footnote{The connection between surface density and acceleration is axiomatic in MOND \citep{milg83}.
Whether it holds in \LCDM\ depends on the specifics of galaxy formation theory.}.
This places them deep in the MOND regime where the details of the interpolation function are irrelevant.

While all dwarfs are in the deep MOND limit, it is less obvious that they qualify as isolated systems.
We check whether the assumption of isolation holds by estimating
the external acceleration $g_{ex}$ upon the dwarf due to M31, and comparing that to the internal acceleration $g_{in}$.
As is the case for the satellites of the Milky Way, $g_{ex}$ is often comparable to $g_{in}$.
In many cases, the two are the same within the errors.

We therefore also report the velocity dispersion for the EFE dominated case in column 7 of Table~\ref{comptable}.
In most cases, the  values for $M_*/L_V = 2\;\MLsun$ are not much different than in the isolated case.
In the deep-MOND limit the ratio of the dispersions calculated for the two cases is $\s_{efe}/\s_{iso} = \sqrt{g_{in}/g_{ex}}$.
Consequently, the case that reports the lower velocity dispersion is always the formally correct one.
We therefore report $\s_{efe}$ in Table~\ref{comptable} only if it is predicted to be less than that in the isolated case for
at least one choice of mass-to-light ratio.  The range in predicted velocity dispersion spanned by the same
range of $1\le M_*/L_V\le 4$ is larger than in the isolated MOND case, owing to the Newtonian $\s \propto M^{1/2}$
scaling\footnote{As in the isolated case, the uncertainty only represents the
range in plausible mass-to-light ratios and not any observational or systematic errors.
There is more opportunity for the latter in the EFE dominated
case since we now need to know the rotation curve of M31 to large distances and the M31-centric distance of each dwarf.}
rather than the MOND $\s \propto M^{1/4}$ scaling. Here too we have not included the uncertainties due to distance and luminosity uncertainties.\footnote{Uncertainties in the distance $D$ from us to the dwarf are generally not so large, since the dwarfs are roughly at the distance of M31.
Since $D$ enters the predicted values of $\s_{iso}$ in a square root, uncertainties in $\s_{iso}$ due to those in $D$ are expected to be small
compared with those due to uncertainties in $M/L$. However, distance uncertainties are much more important for the predicted $\s_{efe}$.
Distance still enters $\s_{efe}$ in a square root [through $(M/r_{1/2})^{1/2}$], but, more importantly, errors in $D$ can be greatly magnified as errors in
there M31-centric distance $D_{M31}$. This enters $\s_{efe}$ in a square root [$\s_{efe}$ scales as $(D_{M31} D)^{1/2}$],
besides entering the isolation criterion itself.  For some dwarfs these uncertainties are quite significant, e.g., for And IX.}

Any one dwarf may transition from the isolated to EFE regime for different choices of mass-to-light ratio.
This is denoted in Fig.~\ref{fig1} by marking each point as a filled or open symbol for isolated or EFE cases, respectively.
One should bear in mind that the assumption of isotropy is immediately suspect in cases dominated by the EFE, which can induce anisotropy.
The EFE predictions are thus not as robust as those for isolated dwarfs, and we tabulate for reference $\s_{iso}$ in all cases, even when
a dwarf appears to be entirely in the EFE regime.

\subsection{Individual Cases}
\label{indcase}

Our predictions are most secure for isolated dwarfs.  Based on the calculation described above, dwarfs that appear to be isolated for all choices
of mass-to-light ratio are And II, And VI, And VII, And XVI, and And XVIII.  And V is also likely to be in this category, though for $M_*/L_V = 1\;\MLsun$
it becomes marginally EFE dominated if we adopt the photometric data of \citet{tollerud2012} rather than those of \citet{collins11}.
Dwarfs that appear to be dominated by the EFE for all choices of
mass-to-light ratio are And IX, And X, And XII, And XIV, and And XXI.  The others span both regimes.  This designation only reflects the assumed variation in
$M_*/L_V$; in many cases the observed internal acceleration $3 \s_{obs}^2/r_{1/2}$ is difficult to distinguish from the estimated external acceleration
$V_{M31}^2/D_{M31}$ given the observational uncertainties.

And V is the worst case among the present data.  It has a predicted velocity dispersion of $\s_{iso} = 5.5\;\kms$
for $M_*/L_V = 2\;\MLsun$ compared to $10.5\pm1.1$ from \citet{tollerud2012} and $11.5^{+5.3}_{-4.4}$ from \citet{collins11}.
In terms of mass-to-light ratios, the observed velocity dispersions imply $M_*/L_V \approx 10\;\MLsun$, implausibly high for a stellar population.
The important question, however, is how significant the difference is.  The stated uncertainty of \citet{tollerud2012}
is considerably smaller than that of \citet{collins11}.  The velocity dispersion we compute for $M_*/L_V = 2\;\MLsun$ is $4.5 \sigma$ smaller
than the measurement of \citet{tollerud2012} but only $1.4 \sigma$ smaller than the measurement of \citet{collins11}.
Given the nature of the observational data, and the way in which the velocity dispersion can change
from observation to observation, it is unclear how much weight to place on the stated uncertainties.

The predicted velocity dispersion is too low in both And XIII and And XVIII, though here the significance is less ($< 2 \sigma$).
In the case of And XIII, there are two independent observations, both higher than predicted albeit by a modest amount ($\sim 1.6\sigma$).
There is only one observed value of the velocity dispersion in And XVIII at present,
so there is no independent indicator of the accuracy of the uncertainty.
Taken at face value, the predicted velocity dispersion is $1.8 \sigma$ too low for $M_*/L_V = 2\;\MLsun$.
These do not seem like particularly worrisome discrepancies given the nature of the data and the necessary assumptions in the prediction.

And IX is an interesting case.  The predicted dispersion is low but consistent with the measurement of \citet{collins10}.  It is substantially lower
than that of \citet{tollerud2012}.  The data differ, so the interpretation depends on which measurement happens to come closer to reality.
However, there is a further complication in the case of And IX.  Of all the dwarfs discussed here, this object is most clearly
subject to strong tides from M31 based on the criteria\footnote{And IX is currently within the MOND-prescribed Roche limit of M31.
How it responds to this strong external perturbation depends on its orbit: how close it comes to M31 and how long it spends so close.}
discussed by \citet{bradadwarf}.  This can result in the induced
anisotropy mentioned in \S~\ref{estimates}.  Depending on its orbit, And IX may also be subject to tidal disruption and dissolution
along the lines inferred for the ultrafaint dwarfs of the Milky Way \citep{mw10}.  Consequently, the assumptions that underpin
our analysis may be violated, making our prediction for And IX the least secure of all the Andromeda dwarfs.

Another interesting case is And II.
\citet{andiicote} made an early velocity dispersion measurement of $9.3^{+2.7}_{-2.6}\;\kms$ that is
in good agreement with the MOND prediction.  This appeared to change with the observations of \citet{kalirai10},
who found a lower velocity dispersion with a much smaller uncertainty:  $7.3\pm0.8\;\kms$.  The MOND prediction
for $M_*/L_V = 2\;\MLsun$ is $4.6\s$ higher than this subsequently measured value.
However, \citet{kalirai10} were only able to measure one side of the dwarf,
leaving open the possibility that the difference in spatial coverage was the cause of the apparent difference in measured velocity dispersions.
Very recently, \citet{Ho2012} have reported substantial rotation in And II, with $v = 10.9\pm2.4\;\kms$ in addition to a dispersion
of  $\sigma = 7.8\pm1.1\;\kms$.  This makes And II unique among dwarf spheroidals in having substantial rotational as well as pressure support.
We incorporate both rotation and dispersion through $\s = \sqrt{\s_{los}^2+\beta v^2}$ with $\beta = 1/3$
since ordered rotation enters the 3D rms velocity of eq.~\ref{i} in quadrature.
This prescription is not exact, as the plane of maximum rotation is unknown, and the maximum rotation speed is not the mass weighted mean.
These two effects tend to counteract one another, so this is probably the best available estimator \citep{boom}.
Certainly it is an improvement on estimates made in ignorance of rotation.
Taking rotation into account, the data are in good agreement with the prediction of MOND.

Indeed, the bulk of the sample is in good agreement with MOND.  The observed and predicted velocity dispersion coincide for
And I, And II, And III, And VI, And VII, And X, And XI, And XII, And XIV, And XV, And XVI, And XXI, and And XXII.
For the cases in the isolated, deep MOND limit where our prediction is most secure, agreement is frequently bang on:
compare $\sigma_{obs}$ with $\s_{iso}$ for objects lacking $\s_{efe}$ in Table~\ref{comptable}.
And XVIII is a bit off, albeit by $< 2 \sigma$ as discussed above.
Only And V disagrees with MOND by as much as $4\s$, and only then for one of the two measurements for that object.
The velocity dispersion reported for And II by \citet{kalirai10} appeared to be problematic,
but the detection of some rotational support in this dwarf by \citet{Ho2012} brings it into agreement with the prediction of MOND.

\subsection{Further Predictions}
\label{furtherpredictions}

\citet{McCLG} provides photometric information for ten dwarfs for which no velocity dispersion is currently available.
We can use this information to predict their velocity dispersions.  MOND predictions
for And XVII, And XIX, And XX, And XXIII, And XXIV, And XXV, And XXVI, And XXVII, And XXVIII, and And XXIX are given in Table~\ref{predptable}
for the range $1\le M_*/L_V\le 4\;\MLsun$.

\begin{deluxetable}{lccccc}
\tablewidth{0pt}
\tablecaption{Predicted Velocity Dispersions}
\tablehead{
\colhead{Dwarf} & \colhead{$L_V$} & \colhead{$r_{1/2}$} & \colhead{$\sigma_{iso}$} & \colhead{$\sigma_{efe}$} & \colhead{$g_{in}$} \\
 & \colhead{$10^5\;L_{\sun}$} & pc & \multicolumn{2}{c}{\kms} &\colhead{$a_0$}
}
\startdata
And XVII   & \phn2.6\phn & \phn381 & $4.5^{+0.9}_{-0.7}$ & $2.5^{+1.0}_{-0.7}$ & 0.043 \\
And XIX    & \phn4.1\phn & 2244 & $5.0^{+1.0}_{-0.8}$ & $2.6^{+1.1}_{-0.8}$ & 0.009 \\
And XX 	& \phn0.28  & \phn165  &  $2.6^{+0.5}_{-0.4}$  & $2.1^{+0.9}_{-0.6}$ & 0.033 \\
And XXIII  & 10.\phn\phn & 1372 & $6.4^{+1.2}_{-1.0}$ & $4.4^{+1.8}_{-1.3}$ & 0.024 \\
And XXIV   & \phn0.94 & \phn489 & $3.5^{+0.7}_{-0.6}$ & $2.8^{+1.2}_{-0.8}$ & 0.020 \\
And XXV    & \phn6.5\phn & \phn945 & $5.7^{+1.1}_{-0.9}$ & $3.5^{+1.5}_{-1.0}$ & 0.027 \\
And XXVI   & \phn0.59 & \phn296 & $3.1^{+0.6}_{-0.5}$ & $2.0^{+0.8}_{-0.6}$ & 0.027 \\
And XXVII  & \phn1.2\phn & \phn579 & $3.7^{+0.7}_{-0.6}$ & $1.8^{+0.7}_{-0.5}$ & 0.020 \\
And XXVIII & \phn2.1\phn & \phn284 & $4.3^{+0.8}_{-0.7}$ & \dots & 0.053 \\
And XXIX   & \phn1.8\phn & \phn481 & $4.1^{+0.8}_{-0.7}$ & $3.8^{+1.6}_{-1.1}$ & 0.028
\enddata
\tablecomments{Dwarfs listed here currently lack published kinematics.  The velocity dispersion predicted by MOND is
based on the photometric data in Table~3 of \citet{McCLG}.
We have taken the 3D $r_{1/2}$ to be 4/3 of the 2D half light radius reported by \citet{McCLG} for consistency with Table~\ref{comptable}.
Columns are the same as in Table~\ref{comptable}.}
\label{predptable}
\end{deluxetable}

These objects are deep in the MOND regime.
Being quite faint, they are predicted to have low velocity dispersions: $\s \lesssim 5 \kms$ for $M_*/L_V = 2\;\MLsun$.
The \textit{a priori} predictions given in Table~\ref{predptable} can be tested as kinematic measurements become available.

\section{Conclusions}
\label{discussion}

MOND appears to be in good agreement with the observed velocity dispersions of the dwarf spheroidals of M31.
Simply looking at Fig.~\ref{fig1}, one might easily mistake the MOND predictions for a set of observational data.

There is no \textit{a priori} guarantee that MOND will work in these systems.  In the context of the dark matter paradigm, we might
observe dwarfs that reside in the $\sim 30\;\kms$ sub-halos commonly seen in \LCDM\ simulations \citep{2big2fail},
or tidally formed dwarfs devoid of dark matter \citep{bournaud} with equilibrium Newtonian velocity dispersions $< 2\;\kms$.
These are plausible scenarios with dark matter that would result in velocity dispersions that MOND could not explain.

Given the photometric data, the velocity dispersion follows in MOND.
The value of $a_0$ is fixed to that established long ago for much brighter rotating galaxies \citep{BBS}.
The only computational flexibility is in the stellar mass-to-light ratio, which is unavoidable in any theory.
For a modest ($\pm$ factor of two) range of $M_*/L$ motivated by our expectations for stellar populations,
the predictions of MOND are immediately in the right vicinity.

As the velocity dispersion data continue to improve, it may be possible to infer $M_*/L$ for individual dwarfs.
We can then ask whether the MOND-inferred mass-to-light ratios correlate with indicators of population age observed in color-magnitude diagrams.
This is already claimed to be the case in the dSph population of the Milky Way \citep{hernandezmond}, and it has been known for some time
that MOND mass-to-light ratios correlate with color in the expected sense in spiral galaxies \citep{SV98,SM02,famaey12}.

The velocity dispersions given in Table~\ref{predptable} are genuine \textit{a priori} predictions subject to observational
confirmation or refutation.  Those listed in Table~\ref{comptable} are also genuine predictions in the weaker sense often used in the
scientific literature.  Though observed velocity dispersions are known for the seventeen dwarfs in Table~\ref{comptable},
these do not inform the computation of the MOND velocity dispersions in any way.
Everything follows from the application of the equations given in \S \ref{estimates} to the photometric data:
the predicted velocity dispersion follows irrespective of the observed one.

\acknowledgements
We thank the referee for constructive comments and for suggesting the consideration of a narrower range of mass-to-light ratios.
%This work of SSM is supported in part by NSF grant AST0908370.

\end{document}